# Easy Impossibility Proofs for $k$-Set Agreement in Message Passing Systems

Martin Biely[*], Peter Robinson[†], and Ulrich Schmid[‡]

[*] EPFL
Switzerland
martin.biely@epfl.ch
[†] Division of Mathematical Sciences, Nanyang Technological University
Singapore
robinson@ntu.edu.sg
[‡] Embedded Computing Systems Group
Technische Universität Wien, Austria
s@ecs.tuwien.ac.at

**Abstract**

Despite of being quite similar agreement problems, consensus and general $k$-set agreement require surprisingly different techniques for proving the impossibility in asynchronous systems with crash failures: Rather than relatively simple bivalence arguments as in the impossibility proof for consensus (= 1-set agreement) in the presence of a single crash failure, known proofs for the impossibility of $k$-set agreement in systems with $f \geqslant k > 1$ crash failures use algebraic topology or a variant of Sperner's Lemma. In this paper, we present a generic theorem for proving the impossibility of $k$-set agreement in various message passing settings, which is based on a simple reduction to the consensus impossibility in a certain subsystem.

We demonstrate the broad applicability of our result by exploring the possibility/impossibility border of $k$-set agreement in several message-passing system models: (i) asynchronous systems with crash failures, (ii) partially synchronous processes with (initial) crash failures, and (iii) asynchronous systems augmented with failure detectors. In (i) and (ii), the impossibility part is just an instantiation of our main theorem, whereas the possibility of achieving $k$-set agreement in (ii) follows by generalizing the consensus algorithm for initial crashes by Fisher, Lynch and Patterson. In (iii), applying our technique yields the exact border for the parameter $k$ where $k$-set agreement is solvable with the failure detector class $(\Sigma_k, \Omega_k)_{1 \leqslant k \leqslant n-1}$ of Bonnet and Raynal. Considering that $\Sigma_k$ was shown to be necessary for solving $k$-set agreement, this result yields new insights on the quest for the weakest failure detector.

This work has been supported by the Austrian Science Foundation (FWF) project P20529 and the Nanyang Technological University grant M58110000.

## I. Introduction

Agreement problems like consensus and set agreement are undoubtly the most prominent target for exploring the solvability/impossibility border in fault-tolerant distributed computing. In such problems, every process $p_i$, $1 \leqslant i \leqslant n$, in a distributed system owns a local proposal value $x_i$, and the problem is to irrevocably compute local output values (also called decision values) $y_i$ that satisfy certain properties. For consensus, no two processes may decide on different values, for set agreement, the number of different decision values must be at most $n-1$ system-wide. An obvious generalization is $k$-set agreement, which requires that the number of different decision values is at most $k$; clearly, consensus is just 1-set agreement, whereas set agreement is equivalent to $(n-1)$-set agreement.

Due to the landmark FLP impossibility result [14], which employs (now classic) combinatorial arguments (bivalence proofs), it is well-known that consensus is impossible to solve in asynchronous systems if a single process may crash. The corresponding result for general $k$-set agreement is the impossibility of solving this problem in asynchronous systems if $f \geqslant k$ processes may crash. Surprisingly, establishing this result requires quite involved techniques based on algebraic topology or a variant of Sperner's lemma [4], [17], [23].

Another very simple and well-known technique for establishing impossibility results are partitioning arguments, which have been used successfully for many distributed computing problems [13]. Essentially, a partitioning argument exploits the fact that one cannot guarantee agreement among those processes of a distributed system that never, neither directly nor indirectly, communicate with each other. In this paper, we use partitioning arguments in a—to the best of our knowledge—new way: as a means for reduction.

More specifically, we present a theorem that provides us with a surprisingly generic tool for proving the impossibility of $k$-set agreement in message-passing systems. It works by reducing the impossibility of $k$-set agreement to the impossibility of achieving consensus in a certain subsystem: In a nutshell, failures and asynchrony in the models considered allow to partition the system into $k$ parts, the processes of which must decide on their own and hence, by choosing distinct proposal values, on different values. Obviously, this leads to at least $k$ different decision values system-wide. The impossibility of $k$-set agreement then follows by showing that it is impossible to reach consensus in at least one of these parts.

**Related work:**

Actually, we are not aware of much research that uses similar ideas: We have employed reduction already in [2] to show that consensus is impossible in certain partially synchronous models, and to prove the tightness of our generalized loneliness failure detector $\mathcal{L}(k)$ for $k$-set agreement. We also learned recently that similar reduction arguments are employed in [5]. In [1], reduction to asynchronous set agreement is used to derive a lower bound on the minimum size of a "synchronous window" that is necessary for $k$-set agreement.

**Detailed Contributions:**

- We present a generic impossibility result for $k$-set agreement that can be applied to a wide variety of message-passing system models and failure assumptions. Our result neither assumes specific assumptions on the (a)synchrony of the model nor on the types of failures that can occur. While the main purpose of this theorem is to derive general impossibility results that hold for *all* algorithms in a specific model, it also turned out to be useful for quickly checking whether a candidate algorithm allows runs that make $k$-set agreement impossible.
- We introduce the notion of $T$-independence for message passing systems, which is related to the progress condition formalism of [24].

- We revisit the impossibility of $k$-set agreement in asynchronous systems with crash failures (some of which are not initial crashes), with (and without) partially synchronous processes. Applying our generic theorem reveals the border that separates impossibility and possibility in this setting.
- Furthermore, by extending the algorithm for initial crashes of [14] to general $k$-set agreement, we show that the impossibility border is tightly matched.
- Finally, we shift our focus to asynchronous systems with failure detectors. We use our theorem to show that $(\Sigma_k, \Omega_k)$ is too weak to solve $k$-set agreement for $1 < k < n-1$.

## II. System Models and Failure Assumptions

We use the computing model of [11], extended with the possibility of querying failure detectors. In [11], 32 different models are defined by varying 5 core system parameters (e.g., synchrony of processes and communication, transmission mechanism, etc.), each of which can be chosen in a way that is either favourable (F) or unfavourable (U) for the algorithm. Informally speaking, we add a $6^{\text{th}}$ dimension to the model:

6. Failure Dectectors
 U. Processes do not have access to failure detectors.
 F. Processes can query a failure detector at the beginning of each step.

For the sake of brevity, we will not repeat the whole formal model of [11] here. Instead, we just introduce the necessary notations and explain the changes necessary for dealing with $k$-set agreement. The details of the case where failure detectors are available will be filled in in Section II-C.

We consider a system $\Pi = \{p_1, \ldots, p_n\}$ of $n$ processes with unique id's $\{1, \ldots, n\}$ that communicate via message-passing, using messages taken from some (possibly infinite) universe $M$.

The communication subsystem is modeled by one buffer per process, which contains messages that have been sent to that process but not yet received. Every process $p \in \Pi$ is modeled as a deterministic state machine, which has a local state (program counter, local variables) that incorporates an input value $x_p$ initialized to some value from a finite set of values $V$, and a write-once output value $y_p \in V \cup \{\bot\}$ initialized to $\bot \notin V$. All other components of the local state are initialized to some fixed value.

State transitions are guided by a transition relation, which atomically takes the current local state of $p$, a (possibly empty) subset of messages $L$ from $p$'s current message buffer, and, in case of failure detectors, a value from the failure detector's domain, and provides a new local state. Sending of messages is guided by a deterministic message sending function, which determines a possibly empty set of messages that are to be sent to the processes in the system, i.e., maps the current state and the subset of messages $L$ to a subset of $\Pi \times M$. Every message $(q, m)$ in this subset is sent by just putting $m$ into $q$'s buffer.

A configuration of the system consists of the vector of local states and the message buffers of all the processes; in the initial configuration, all processes are in an initial state and the message buffers are empty.

A run $\rho = (C_0, C_1, \ldots)$ is an infinite sequence of configurations that starts from an initial configuration $C_0$, and $C_{i+1}$ results from a legitimate (according to the transition relation and message sending function) step of a single process $p$ in configuration $C_i$.

The above basic model is strengthened by restricting the set of runs by some admissibility conditions that depend on the particular system model $\mathcal{M}$ used. For example, the FLP model [14], denoted as $\mathcal{M}^{\text{ASYNC}}$, requires that (1) every correct process takes an infinite number of steps, (2) faulty processes execute only finitely many steps and may omit sending messages to a subset of receivers in the very last step, and (3) every message sent by a process to a correct receiver process is eventually received.

With the exception of Section III, we will assume systems adhering to the asynchronous model $\mathcal{M}^{\text{ASYNC}}$, sometimes augmented with a failure detector (Section II-C) or with the assumption of partially synchronous processes (Section V).

## A. $k$-Set Agreement

We study distributed algorithms that solve agreement problems, namely, $k$-set agreement. Their purpose is to compute and irrevocably set the output $y_p$ of process $p$ to some decision value, based on the proposal values $x_q \in V$, for $1 \leqslant q \leqslant n$ and $|V| \geqslant n$,[1] which must satisfy the following properties:

**$k$-Agreement:** Processes must decide on at most $k$ different values.
**Validity:** If a process decides on $v$, then $v$ was proposed by some process.
**Termination:** Every correct process must eventually decide.

Note that the agreement property binds together the decision values of all (correct or faulty) processes. For $k = 1$, $k$-set agreement is hence equivalent to uniform consensus [7]. It follows from [14] that non-uniform and hence also uniform consensus cannot be solved in asynchronous systems if just one process may crash.

## B. Restrictions of Algorithms and Indistinguishability of Runs

We will occasionally use a subsystem $\mathcal{M}'$ that is a *restriction* of $\mathcal{M}$, in the sense that it consists of a subset of processes in $\Pi$, while using the same mode of computation (atomicity of computing steps, time-driven vs. message-driven, etc.) as $\mathcal{M}$. We make this explicit by using the notation

$$\mathcal{M} = \langle \Pi \rangle \text{ and } \mathcal{M}' = \langle D \rangle,$$

for some set of processes $D \subseteq \Pi$. Note that this definition does not imply anything about the synchrony assumptions which hold in $\mathcal{M}'$. All that is required is that $\mathcal{M}'$ is computationally compatible with $\mathcal{M}$: Any algorithm designed for $\mathcal{M}$ can also be run in $\mathcal{M}'$, albeit on a smaller set of processes.

**Definition 1** (Restriction of an Algorithm). Let $A$ be an algorithm that works in system $\mathcal{M} = \langle \Pi \rangle$ and let $D \subseteq \Pi$ be a nonempty set of processes. Consider a restricted system $\mathcal{M}' = \langle D \rangle$. The *restricted algorithm* $A_{|D}$ for system $\mathcal{M}'$ is constructed by dropping all messages sent to processes outside $D$ in the message sending function of $A$, obtaining the message sending function of $A_{|D}$.

Note that we do not change the actual code of algorithm $A$ in any way. In particular, the restricted algorithm still uses the value of $|\Pi|$ for the size of the system, even though the real size of $D$ might be much smaller.

Whereas this is sufficient for running an algorithm designed for $\mathcal{M}$ in the restricted system $\mathcal{M}'$, in practice, one would also remove any dead code (resulting from state transitions triggered by message arrivals from processes in $\Pi \setminus D$, from the transition relation of $A$ to obtain the actual transition relation of $A_{|D}$. Note that we use $\mathcal{M}_A$ to denote the set of runs of algorithm $A$ in model $\mathcal{M}$.

We will use a concept of similarity/indistinguishability of runs that is slightly weaker than the usual notion [20, Page 21], as we require the same states only *until* a decision state is reached. This makes a difference for algorithms where $p$ can help others in reaching their decision after $p$ has decided, for example, by forwarding messages.

**Definition 2** (Indistinguishability of Runs). Two runs $\alpha$ *and* $\beta$ *are indistinguishable (until decision) for a process* $p$, if $p$ has the same sequence of states in $\alpha$ and $\beta$ until $p$ decides. By $\alpha \overset{D}{\sim} \beta$ we denote the fact that $\alpha$ and $\beta$ are indistinguishable (until decision) for every $p \in D$.

---

[1] The assumption $|V| \geqslant n$ allows runs where all processes start with different propose values.

**Definition 3** (Compatibility of Runs). Let $\mathcal{R}$ and $\mathcal{R}'$ be sets of runs. We say that *runs $\mathcal{R}'$ are compatible with runs $\mathcal{R}$ for processes in $D$*, denoted by $\mathcal{R}' \preccurlyeq_D \mathcal{R}$, if $\forall \alpha \in \mathcal{R}' \; \exists \beta \in \mathcal{R} \colon \alpha \overset{D}{\sim} \beta$.

## C. Failure Detectors and Failure Patterns

A failure detector [6] $\mathcal{D}$ is an oracle that can be queried by processes in any step, before making a state transition. The definition of failure detectors is based on the notion of a global *time*, which we did not introduce yet. Recall that a run is a sequence of configurations, where $C_i$ results from a single step of a single process $p$ in configuration $C_{i-1}$. We call this step the $i$th step of the run, and consider it to occur at time $i$. Note that processes do not have access to time.

The *failure pattern* $F(.)$ of a run $\alpha$ is a function that outputs the set of crashed processes for a given time $t$; that is, $p \in F(t)$, if there is no $i \geqslant t$, such that the $i$th step of the run is a step of $p$. Moreover, we denote the *set of faulty processes* in the run as $F = \bigcup_{t \geqslant 0} F(t)$.

The behaviour of $\mathcal{D}$ in a run $\alpha$ depends on the failure pattern $F(.)$, which defines the set of admissible *failure detector histories*. The value of a query of a process $p$ in a step at time $t$ is defined by the *history function* $H(p, t)$, which maps process identifiers and time to the *range* of output symbols of $\mathcal{D}$. Clearly, a run in a system augmented with failure detectors is *admissible*, if all state transitions occur according to a legal history $H$ of $\mathcal{D}$, given the failure pattern of the run.

We denote the augmented asynchronous model, where runs are admissible in $\mathcal{M}^{\text{ASYNC}}$ and processes can query failure detector $\mathcal{D}$ in any step, as $\langle \mathcal{M}^{\text{ASYNC}}, \mathcal{D} \rangle$. If there is an algorithm $A$ that solves problem $P$ in $\langle \mathcal{M}^{\text{ASYNC}}, \mathcal{D} \rangle$, we say that $\mathcal{D}$ *solves* $P$. We say that algorithm $A_{\mathcal{D} \to \mathcal{D}'}$ transforms $\mathcal{D}$ to $\mathcal{D}'$, if processes maintain output variables $output_{\mathcal{D}'}$ that emulate failure detector histories of $\mathcal{D}'$, which are admissible for $F(.)$.

Based on this notion of transforming oracles, [6] introduces a comparison relation on failure detectors: We say that $\mathcal{D}'$ is *weaker* than $\mathcal{D}$ and call $\mathcal{D}$ *stronger* than $\mathcal{D}'$, if such an algorithm $A_{\mathcal{D} \to \mathcal{D}'}$ exist. If there is also an algorithm $A_{\mathcal{D}' \to \mathcal{D}}$, we say that $\mathcal{D}$ and $\mathcal{D}'$ are *equivalent*. If no such algorithm $A_{\mathcal{D}' \to \mathcal{D}}$ exists, we say that $\mathcal{D}$ is *strictly stronger* than $\mathcal{D}'$; *strictly weaker* is defined analogously. If neither $A_{\mathcal{D} \to \mathcal{D}'}$ nor $A_{\mathcal{D}' \to \mathcal{D}}$ exists then we say that $\mathcal{D}$ and $\mathcal{D}'$ are *incomparable*. A failure detector $\mathcal{D}'$ is the *weakest for problem $P$* if $\mathcal{D}$ is weaker than any failure detector $\mathcal{D}$ that solves $P$.

While the weakest failure detector for message passing $k$-set agreement is still unknown, the *quorum family* $\Sigma_k$ was shown in [3] to be necessary for solving $k$-set agreement with any failure detector $\mathcal{X}$, in the sense that there is a transformation that implements $\Sigma_k$ in the system $\langle \mathcal{M}^{\text{ASYNC}}, \mathcal{X} \rangle$.

We will now restate the failure detector classes $\Sigma_k$ and $\Omega_k$; see [22] for a recent overview of failure detectors for $k$-set agreement.

**Definition 4** (cf. [3]). The *generalized quorum failure detector* $\Sigma_k$, with $\Sigma = \Sigma_1$, outputs a set of trusted process ids, such that for all environments $\mathcal{E}$ and for all failure patterns $F(.) \in \mathcal{E}$ the following holds:

*Intersection:* For every set of $k+1$ processes $\{p_1, \ldots, p_{k+1}\}$ and for all $k+1$ time instants $t_1, \ldots, t_{k+1}$, there exist indices $i$ and $j$ with $1 \leqslant i \neq j \leqslant k+1$, such that $H(p_i, t_i) \cap H(p_j, t_j) \neq \emptyset$.
*Liveness:* $\exists t \; \forall t' \geqslant t \; \forall p_i \notin F \colon H(p_i, t') \cap F = \emptyset$.

If a process $p$ crashes at time $t$, i.e., $p \in F(t)$, we define $\forall t' \geqslant t \colon H(p, t') = \Pi$.

**Definition 5** (cf. [21]). The output of the *generalized leader oracle* $\Omega_k$, for $1 \leqslant k \leqslant n-1$, satisfies the following properties:

*Validity:* For all processes $p$ and all times $t$, history $H(p, t)$ is a set of $k$ process identifiers.
*Eventual Leadership:* There exists a time $t_{\text{GST}}$ and a set $LD$, such that

$$(LD \cap (\Pi \setminus F) \neq \emptyset) \wedge (\forall t \geqslant t_{\text{GST}} \; \forall p \colon H(p, t) = LD).$$

## III. The Impossibility Theorem

In this section, we will present our general $k$-set agreement impossibility theorem. Due to its very broad applicability, the theorem itself is stated in a highly generic and somewhat abstract way. It captures a reasonably simple idea, however, which boils down to extracting a consensus algorithm for a certain subsystem where consensus is unsolvable: Suppose that a given $k$-set agreement algorithm $A$ for some system model $\mathcal{M}$ has runs, where processes start with distinct values and $k$ partitions $D_1, \ldots, D_{k-1}$ and $\overline{D}$ can be formed: Processes in the $k-1$ partitions $D_i$ decide on (at least) $k-1$ different values, and no process in partition $\overline{D}$ ever hears from any process in $D_i$ before it decides. Note carefully that processes in $\overline{D}$ can communicate arbitrarily within $\overline{D}$. Then, the ability of $A$ to solve $k$-set agreement would imply that the restricted algorithm $A_{|\overline{D}}$ can solve consensus in the restricted model $\mathcal{M}' = \langle \overline{D} \rangle$. However, if the synchrony and failure assumptions of $\mathcal{M}$ are such that consensus cannot be solved in $\mathcal{M}'$, this is a contradiction. This intuition will become completely clear when we apply Theorem 1 in Sections V and VII.

**Theorem 1** ($k$-Set Agreement Impossibility). *Let $\mathcal{M} = \langle \Pi \rangle$ be a system model and consider the runs $\mathcal{M}_A$ that are generated by some fixed $k$-set agreement algorithm $A$ in $\mathcal{M}$, where every process starts with a distinct input value. Fix some nonempty disjoint sets of processes $D_1, \ldots, D_{k-1}$, and a set of distinct decision values $\{v_1, \ldots, v_{k-1}\}$. Moreover, let $D = \bigcup_{1 \leqslant i < k} D_i$ and $\overline{D} = \Pi \setminus D$. Consider the following two properties:*

**(dec-$D$)** *For every set $D_i$, value $v_i$ was proposed by some process in $D$, and there is some process in $D_i$ that decides on $v_i$.*

**(dec-$\overline{D}$)** *If $p_j \in \overline{D}$ then $p_j$ receives no messages from any process in $D$ until after every process in $\overline{D}$ has decided.*

*Let $\mathcal{R}_{(\overline{D})} \subseteq \mathcal{M}_A$ and $\mathcal{R}_{(D,\overline{D})} \subseteq \mathcal{M}_A$ be the sets of runs of $A$ where (dec-$\overline{D}$) respectively both, (dec-$D$) and (dec-$\overline{D}$), hold.[2] Suppose that the following conditions are satisfied:*

**(A)** $\mathcal{R}_{(\overline{D})}$ *is nonempty.*
**(B)** $\mathcal{R}_{(\overline{D})} \preccurlyeq_{\overline{D}} \mathcal{R}_{(D,\overline{D})}$.

*In addition, consider a restricted model $\mathcal{M}' = \langle \overline{D} \rangle$ such that the following hold:*

**(C)** *There is no algorithm that solves consensus in $\mathcal{M}'$.*
**(D)** $\mathcal{M}'_{A_{|\overline{D}}} \preccurlyeq_{\overline{D}} \mathcal{M}_A$.

*Then, $A$ does not solve $k$-set agreement in $\mathcal{M}$.*

*Proof:* For the sake of a contradiction, assume that there is a $k$-set agreement algorithm $A$ for model $\mathcal{M}$, sets of runs $\mathcal{R}_{(\overline{D})}$ and $\mathcal{R}_{(D,\overline{D})}$ and some sets of processes $D_1, \ldots, D_{k-1}$ such that conditions (A)–(D) hold. Due to (A) we have $\mathcal{R}_{(\overline{D})} \neq \emptyset$; then, (B) implies that $\mathcal{R}_{(D,\overline{D})}$ is nonempty too. Observe that (dec-$D$) ensures that there are $\geqslant k-1$ distinct decision values among the processes in $D$, in every run in $\mathcal{R}_{(D,\overline{D})}$. Since algorithm $A$ satisfies $k$-agreement, the compatibility requirement (B) between runs $\mathcal{R}_{(\overline{D})}$ and $\mathcal{R}_{(D,\overline{D})}$ for processes in $\overline{D}$ implies the following constraint:

**(Fact 1)** *In each run in $\mathcal{R}_{(\overline{D})}$, all processes in $\overline{D}$ must decide on a common value.*

We will now show that this fact yields a contradiction. Starting from $\mathcal{M}'_{A_{|\overline{D}}}$, i.e., the set of runs of the restricted algorithm in model $\mathcal{M}'$, we know by (D) that for each $\rho' \in \mathcal{M}'_{A_{|\overline{D}}}$, there exists a run $\rho \in \mathcal{M}_A$ such that $\rho' \overset{\overline{D}}{\sim} \rho$. Obviously, no process $p \in \overline{D}$ receives messages from a process $q \in D$ in $\rho'$ before $p$'s decision, as such a process $q$ does not exist in the restricted model $\mathcal{M}'$. Clearly, the same is true for the indistinguishable run $\rho$ (even though such a process $q$ does exist in model $\mathcal{M}$). Therefore, we have that, in fact, $\rho \in \mathcal{R}_{(\overline{D})}$, and due to (Fact 1), we know that in each run $\rho' \in \mathcal{M}'_{A_{|\overline{D}}}$ all processes decide on the same value. This, however, means that we could employ $A_{|\overline{D}}$ to solve consensus in $\mathcal{M}'$, which is a contradiction to (C). ∎

---
[2] Note that $\mathcal{R}_{(\overline{D})}$ is by definition compatible with the runs of the restricted algorithm $A_{|\overline{D}}$.

**Remarks**

There are several noteworthy points about Theorem 1:
- The proof neither restricts the types of failures that can occur in $\mathcal{M}$ nor the underlying synchrony assumptions of $\mathcal{M}$ in any way.
- Our impossibility argument uses a 2-partitioning argument but does *not* require the system to (temporarily or permanently) decompose into $k + 1$ partitions. In particular, there is no further restriction on the communication among processes within $D$ and within $\overline{D}$.
- Despite its main purpose of showing impossibilities, our theorem is also useful when developing new algorithms for achieving $k$-set agreement. For example, suppose that we are given some unproven but seemingly promising new algorithm $A$ for a model close to asynchrony. Then, checking whether the runs of $A$ are such that the conditions of Theorem 1 are satisfied will allow us to determine already at an early stage (i.e., before developing a detailed correctness analysis) whether it is worthwhile to explore $A$ further. In particular, if (dec-$D$) can be satisfied in some runs, i.e., (A) holds, the algorithms is very likely flawed, as the remaining conditions are typically easy to construct in sufficiently asynchronous systems.
- At a first glance, requirement (B) might appear to be redundant. After all, it should always be possible to find a run in $\mathcal{R}_{(D,\overline{D})}$ that is indistinguishable for the processes in $\overline{D}$, given some run in $\mathcal{R}_{(\overline{D})}$. We will now try to give an intuition for its necessity; in the proof of Theorem 10, we will see that (B) is non-trivial in realistic settings.

  To see why (B) is necessary, first consider some run $\gamma$ (of some algorithm in some model $\mathcal{M}$) that satisfies property (dec-$D$). This stipulates $k - 1$ distinct decision values among the processes in $D$, which essentially means that $\gamma$ was a quite "asynchronous" run for the processes in $D$. It could therefore be the case that the synchrony assumptions of $\mathcal{M}$ require $\gamma$ to be "synchronous" for the processes in $\overline{D}$. Now suppose that we are given a run $\alpha \in \mathcal{R}_{(\overline{D})}$ and we need to find a run $\beta \in \mathcal{R}_{(D,\overline{D})}$ that is indistinguishable for processes in $\overline{D}$, in order to make (B) hold. If $\alpha$ is an "asynchronous" run for the processes in $\overline{D}$, we might not be able to find a matching run $\beta \in \mathcal{R}_{(D,\overline{D})}$, as the above setting requires such runs to be "synchronous" for the processes in $\overline{D}$. Consider, for example, the (highly artificial) model where computing speed and communication among processes in $\overline{D}$ is synchronous in a run if and only if the processes in $D$ decide on at least $k - 1$ distinct values. Clearly it does not hold that $\mathcal{R}_{(\overline{D})} \preccurlyeq_{\overline{D}} \mathcal{R}_{(D,\overline{D})}$ in this scenario.

## IV. $T$-Independence

We proceed with introducing a convenient notion for message passing systems, which is similar to the *progress conditions* of concurrent objects [18], [24] in shared memory models. Bear in mind that we only consider algorithms for decision tasks, like $k$-set agreement; that is, every correct process must eventually decide.

**Definition 6** ($T$-independence). Consider a model $\mathcal{M} = \langle \Pi \rangle$ and let $T \subseteq 2^\Pi$ be a family of sets of processes. We say that *A satisfies $T$-independence in* $\mathcal{M}$, if for all sets $S \in T$ it holds that the subset of runs of $A$ in $\mathcal{M}$ where processes in $S$ only receive messages from other processes in $S$ until every process in $S$ either decides or crashes, is nonempty.

If, in addition, the subset of runs of $A$ in $\mathcal{M}$, where processes in $S$ *eventually* only receive messages from other processes in $S$ until every process in $S$ either decides or crashes, is nonempty, we say that $A$ *satisfies strong $T$-independence in $\mathcal{M}$*.

**Observation 1.** *The following properties obviously hold:*
*(a) If algorithm $A$ satisfies strong $T$-independence in $\mathcal{M}$, then $A$ also satisfies $T$-independence in $\mathcal{M}$.*
*(b) If algorithm $A$ satisfies $T$-independence in $\mathcal{M}$ and $T' \subseteq T$, then $A$ satisfies $T'$-independence in $\mathcal{M}$.*

We can express the following classic progress conditions in terms of $T$-independence: *Wait-freedom* [16] provides strong $2^\Pi$-independence. Moreover, *obstruction-freedom* implies $\{\{p_1\}, \ldots, \{p_n\}\}$-independence. The classic assumption of an $f$-resilient algorithm guarantees strong $\{S \mid (S \subseteq \Pi) \wedge (|S| \geq |\Pi| - f)\}$-independence, whereas using (non-strong) $\{S \mid (S \subseteq \Pi) \wedge (|S| \geq |\Pi| - f)\}$-independence holds when up to $f$ *initial crash failures* can be tolerated. Analogously to [18], $T$-independence also enables us to specify *asymmetric progress conditions*, e.g., strong $\{S \mid \{p_1\} \subseteq S \subseteq \Pi\}$-independence is guaranteed by wait-freedom of process $p_1$.

## V. Impossibility in the Partially Synchronous and Asynchronous Case

It is easy to show that $k$-set agreement is impossible in the purely asynchronous model, if we assume a wait-free environment: It suffices to simply delay all communication until every process has decided on its own propose value. When the number of failures is somewhat restricted and/or the model is partially synchronous, however, a more involved argument is necessary. In this section, we will show how to avoid proving the impossibility "from scratch" by instantiating Theorem 1.

**Theorem 2.** *There is no algorithm that solves $k$-set agreement in a system $\mathcal{M}$ of $n$ processes where*
- *processes are synchronous,*
- *communication is asynchronous,*
- *a process can broadcast a message in an atomic step, and*
- *receiving and sending are part of the same atomic step,*

*for any*
$$k \leq \frac{n-1}{n-f}, \tag{1}$$

*even if, of the $f$ possibly faulty processes, $f - 1$ can fail by crashing initially and only one process can crash during the execution.*

*Proof:* Assume in contradiction that some $f$-resilient algorithm $A$ solves $k$-set agreement. We will show that conditions (A)–(D) of Theorem 1 are satisfied, thus yielding a contradiction.

As a first step, we will identify suitable sets $D_i$ such that (A)–(B) hold for the runs in $\mathcal{R}_{(\overline{D})}$ and $\mathcal{R}_{(D,\overline{D})}$, respectively. Let $\ell = n - f$; for $1 \leq i < k$, define $D_i = \{p_{(i-1)\ell+1}, \ldots, p_{i\ell}\}$ and let
$$D = \bigcup_{1 \leq i \leq k-1} D_i.$$

Note that the failure assumption (1) guarantees that these sets $D_i$ exist.

**Lemma 3.** *The set $\overline{D}$ contains at least $n - f + 1$ processes, and every $D_i$, $1 \leq i < k$, contains exactly $\ell = n - f$ processes.*

*Proof:* Since obviously $|D_i| = \ell$, we are done if we can show that $|D| + n - f + 1 \leq n$, i.e.,

$$(k-1)(n-f) + (n-f+1) = k(n-f) + 1 = k\ell + 1 \leq n,$$

which matches exactly (1). ∎

Moreover, the failure bound (1) together with the fact that communication is asynchronous, immediately implies the following lemma:

**Lemma 4.** *Algorithm $A$ is $\{D_1, \ldots, D_{k-1}, \overline{D}\}$-independent.*

We now show the conditions of Theorem 1:
**(A)** By Observation 1.(b) and Lemma 4 it follows immediately that $\mathcal{R}_{(\overline{D})} \neq \emptyset$.
**(B)** Consider the set of runs $\mathcal{H}$ where all communication between the sets of processes $D_1, \ldots, D_{k-1}, \overline{D}$ is delayed until every correct process has decided; Lemma 4 implies that $\mathcal{H} \neq \emptyset$. For any $\rho \in \mathcal{R}_{\overline{D}}$ it is easy to find one $\rho' \in \mathcal{H}$, where all processes in $\overline{D}$ go through the same states until deciding. Moreover, clearly $\mathcal{H} \subseteq \mathcal{R}_{(D, \overline{D})}$, thus establishing (B).
**(C)** Now consider a system $\mathcal{M}' = \langle \overline{D} \rangle$ that has the same system assumptions as $\mathcal{M}$, with the restriction that at most one process can crash in $\mathcal{M}'$ at any time. Condition (C) follows immediately from the result of [11, Table I], since we have already shown in Lemma 3 that

$$|\overline{D}| \geq n - f + 1 \geq 2$$

and one process can crash in the runs of $\mathcal{M}'$.
**(D)** We will show that for every run $\rho' \in \mathcal{M}'_{A_{|\overline{D}}}$, there is a corresponding run $\rho \in \mathcal{M}_A$ such that $\rho' \overset{\overline{D}}{\sim} \rho$. Fix any $\rho' \in \mathcal{M}'_{A_{|\overline{D}}}$ and consider the run $\rho \in \mathcal{M}_A$ where every correct process in $\overline{D}$ has the same sequence of states in $\rho$ as in $\rho'$, and all remaining processes—of which there are $\leq f - 1$—are initially dead in $\rho$. Such a run $\rho$ exists, since $A_{|\overline{D}}$ is the restriction of $A$ (see Definition 1).

We can therefore apply Theorem 1 and conclude that $A$ does not solve $k$-set agreement. ∎

Since an impossibility under stronger assumptions implies impossibility under weaker ones, we have the following corollary:

**Corollary 5.** *The impossibility of $k$-set agreement from Theorem 2 continues to hold under weaker assumptions, in particular, if processes are asynchronous, broadcasts are not possible in one step, sending and receiving within one atomic step is not possible, and all $f$ processes may fail by crashing.*

## VI. Possibility of $k$-Set Agreement with Initially Dead Processes

In this section, we will show that Theorem 2 tightly captures the impossibility of $k$-set agreement, by presenting a matching bound for the solvability of $k$-set agreement in asynchronous systems with $f$ initial crashes.

For the consensus case $k = 1$, we know from [14] that it is sufficient for a majority of processes to be correct. The protocol of [14] operates in two stages: In the first stage, each process broadcasts a message (containing its process id). Every process then waits until it has received $L - 1$ (where $L$ is $\lceil (n+1)/2 \rceil$) messages.

In the second stage, every process broadcasts a message containing its initial value and the list of $L - 1$ processes it has received messages from in the first stage. Then it waits for messages from those $L - 1$ processes it has received messages from in the first stage, and for a message from every remote process mentioned in one of the lists it receives.

Now consider a directed simple graph, in which each node corresponds to a process and there is an edge from $u$ to $w$ iff the process corresponding to $w$ has received a message from the process corresponding to $u$ in the first stage. Let us call this graph $G$. Clearly, every node in $G$ has in-degree $L-1$. Processes only know some part of $G$ after the first stage, but have got complete and consistent knowledge of $G$ after the second stage. At this point, every process can thus consistently determine an initial clique $C$ in $G$, i.e., a fully connected maximal subgraph with no incoming edges. Since $n > 2f$, exactly one such $C$ must exist. A deterministic rule for choosing one of the proposal values of the processes in $C$ (e.g., the value proposed by the process whose identifier is minimal in the clique) is used as the decision value of every process.

For the general case $k \geqslant 1$, we can use the same algorithm if we can make sure that each process can determine one of at most $k$ initial cliques. We will now determine a value for $L$, which guarantees this for some given $k$. Note that the ability to select a value for $L$ is also restricted by $f$. Thus, by combining the relations between $L$ and $k$ and $f$, respectively, we will be able to determine the range of $f$ for which $k$-set agreement is solvable.

We call a strongly connected component $C$ of a directed graph a *source component*, if, in the directed acyclic graph (DAG) generated by contracting all vertices of the strongly connected components of $G$ into single vertices, the vertex corresponding to $C$ is a source, i.e., has in-degree 0.

**Lemma 6.** *Every finite directed simple graph $G = (V, E)$, where each vertex $v$ has at least in-degree $\delta > 0$, has a source component $C$ of size at least $\delta + 1$.*

*Proof:* Obviously, the graph $G'$ obtained from $G$ by contracting all vertices in each strongly connected component is a directed acyclic graph. Like every DAG, $G'$ has at least one vertex $c'$ with in-degree 0. Let $C$ be the set of processes in $G$ that were contracted to $c'$. By definition, $C$ must be a source component, so it remains to show that $|C| \geqslant \delta + 1$. Take any vertex $v \in C$. Clearly, all in-neighbours of $v$ must also be in $C$, since $C$ is a source component. Thus, $C$ must contain at least $\delta$ vertices besides $v$. ∎

**Lemma 7.** *Consider a finite directed simple graph $G$, where each process has at least in-degree $\delta > 0$. In each weakly connected component of $G$, there exists at least one source component $C$ of size at least $\delta + 1$.*

*Proof:* Follows by using the previous theorem for each sub-graph corresponding to a weakly connected component. ∎

From this lemma, it follows that every process has (at least) one directed incoming path from all the processes in (at least) one source component. Moreover, it is easy to see that, when $2\delta \geqslant n$, then there can be only one source component, i.e., no more than $\lfloor n/(\delta+1) \rfloor$.

Returning to the algorithm from [14], we find that detecting locally which processes belong to the initial clique $C$ in $G$ is equivalent to locally detecting which processes belong to the source component a process is connected to. Moreover, as mentioned earlier, waiting for $L-1$ messages in the first stage clearly induces a graph $G$ with $\delta = L-1$, and thus at most $\lfloor n/L \rfloor$ source components. From this it follows that processes will decide on at most $\lfloor n/L \rfloor$ values, so $k$-set agreement with $k \geqslant \lfloor n/L \rfloor$ is indeed solvable.

As our last step, we have to relate $L$ to the bound on the number of initially crashed processes $f$. On one hand, we want $L$ to be as large as possible in order to decrease the number of source components. On the other hand, since processes wait until they have received a message from $L-1$ remote processes in the first stage, it is clearly not advisable to choose $L - 1 \geqslant n - f$. Therefore, we now fix $L = n - f$, which leads to $k$-set agreement being solvable when $k \geqslant \lfloor n/(n-f) \rfloor$. Since $n$, $f$, and $k$ are all integers, we get that $k + 1 > n/(n - f)$ and hence $kn > (k + 1)f$. Note that, for $k = 1$, this matches the requirement of a majority of correct processes.

Considering the border case $kn = (k + 1)f$, we get $n - f = n/(k + 1)$. A standard partitioning argument reveals that $k$-set agreement is impossible in this case: Assume that there is an algorithm $A$ that solves $k$-set agreement in such a system. The above condition on $n$ and $f$ implies that we can partition the system into $k + 1$ disjoint groups of processes $\Pi_0, \ldots, \Pi_k$. From the set of possible input values $V$, choose any $v_0, \ldots, v_k$, s.t., $v_i = v_j \Leftrightarrow i = j$. Clearly, for each $i$, there is an execution $\varepsilon_i$ of $A$ where all processes in $\Pi_i$ have initial value $v_i$ and all processes in $\Pi \setminus \Pi_i$ are initially dead. Since $A$ solves $k$-set agreement, all processes in $\Pi_i$ have to eventually decide on $v_i$ in $\varepsilon_i$. Therefore, by delaying messages between the partitions $\Pi_i$ sufficiently long, it is easy to construct an execution $\varepsilon$ without any initial crashes, which is indistinguishable (until decision) for all $p \in \Pi_i$ from $\varepsilon_i$, $0 \leqslant i \leqslant k$. But now we have $k + 1$ different decision values (i.e., $v_0, \ldots, v_k$) in $\varepsilon$, which contradicts the assumption that $A$ solves $k$-set agreement. Therefore, we have obtained the following result:

**Theorem 8.** *In an asynchronous system with $n$ processes up to $f$ of which may be initially dead, $k$-set agreement is solvable if and only if $kn > (k + 1)f$ or, equivalently,*

$$k > \frac{f}{n - f}.$$

## VII. Impossibility with Failure Detector $(\Sigma_k, \Omega_k)$

In this section, we will demonstrate the full power of Theorem 1 by deriving a new result: We prove the impossibility of achieving $k$-set agreement with failure detector $(\Sigma_k, \Omega_k)$, for all $1 < k < n - 1$. In [5, Theorem 2], it was shown that $k$-set agreement is impossible with $(\Sigma_k, \Omega_k)$ if $1 < 2k^2 \leqslant n$, which is a much more restrictive bound than the one given by Theorem 10 below.

For our impossibility proof, we will make use of a certain stronger failure detector that nevertheless allows up to $k$ partitions. Note that this actually strengthens our impossibility result.

**Definition 7.** *Let $\{D_1, \ldots, D_{k-1}, D_k\}$ be a partitioning of the processes in $\Pi$, and let $\overline{D} = D_k$. The partition failure detector $(\Sigma'_k, \Omega'_k)$ provides failure detector histories with the following properties:*

1) For $1 \leqslant i \leqslant k$, the output of $\Sigma'_k$ at every process in $D_i$ is a valid history for $\Sigma$ $(= \Sigma_1)$ in the restricted model $\mathcal{M}_i = \langle D_i \rangle$ (where only processes from $D_i$ are ever output by $\Sigma$), with an additional condition: Let $t_j$ be the earliest point in time when $p_j \in F(t_j)$, for any $p_j \in D_i$. If $t_j$ is finite, then $\forall t \geqslant t_j$ it holds that the output of $\Sigma'_k$ at $p_j$ is defined to be the whole set $\Pi$.
2) Let $LD^t_j$ denote the set of $k$ leader candidates that is the output of $\Omega'_k$ at process $p_j$ at some point in time $t$. We just assume[3] $\Omega'_k = \Omega_k$, i.e., $LD^t_j$ must satisfy Definition 5 for some stabilization time $t_{\text{GST}}$: There exists a set $LD$ of $k$ processes and an index $1 \leqslant j \leqslant k$ with $LD \cap D_j \cap (\Pi \setminus F) \neq \emptyset$, such that the output $LD^t_j$ of every correct process $p_j \in \Pi$ is $LD$ for all $t \geqslant t_{\text{GST}}$.

We call a history of $(\Sigma'_k, \Omega'_k)$ a *partitioning history*.

**Lemma 9.** *Failure detector $(\Sigma_k, \Omega_k)$ is weaker than $(\Sigma'_k, \Omega'_k)$.*

*Proof:* Consider an arbitrary finite stabilization time $t_{\text{GST}}$. Since $\Omega'_k = \Omega_k$, every history of $(\Sigma'_k, \Omega'_k)$ obviously satisfies the (Eventual Leadership) property of $\Omega_k$.

To show that every history of $(\Sigma'_k, \Omega'_k)$ also satisfies the properties of $\Sigma_k$, choose any set $P$ of $k + 1$ processes in $\Pi$. First, observe that the combined liveness conditions of the local $\Sigma$ histories immediately imply that liveness holds for $\Sigma_k$ (see Definition 4). By the pigeon hole principle, at least two processes of $P$ must be in in the same set $D_i$, for some $1 \leqslant i \leqslant k$, where $D_k = \overline{D}$. Hence, the intersection property of $\Sigma$ in $\langle D_i \rangle$ implies that the history is valid for $\Sigma_k$, which completes the proof. ∎

---

[3]Actually, it would be possible to also strengthen $\Omega'_k$. As this somewhat obfuscates key ideas of the proof, however, we dropped this generalization in this paper.

We are now ready for stating our major theorem:

**Theorem 10.** *There is no $(n-1)$-resilient algorithm that solves $k$-set agreement in an asynchronous system with failure detector $(\Sigma_k, \Omega_k)$, for all $2 \leqslant k \leqslant n-2$.*

*Proof:* We assume by contradiction that there is such an algorithm $A$. Note that there are exactly $n = k - 1 + j$ processes in the system, for some $j \geqslant 3$. Consider the following partitioning of $\Pi$: Let $\overline{D} = \{p_1, \ldots, p_j\}$ and choose $D_1, \ldots, D_{k-1}$ such that they partition the set $\Pi \setminus \overline{D}$; since

$$D = \bigcup_{1 \leqslant i < k} D_i, \text{ i.e., } |D| = n - j = k - 1,$$

such a partitioning exists.

We will actually prove the impossibility for $A$ provided with the stronger failure detector $(\Sigma'_k, \Omega'_k)$. This impossibility can be carried over to $A$ provided with $(\Sigma_k, \Omega_k)$ by using Lemma 9.

We start with two technical lemmas, which justify why we call histories of $(\Sigma'_k, \Omega'_k)$ "partitioning histories": Intuitively speaking, it is straightforward to combine histories at different processes. The first lemma proves that we can "paste together" different executions at partition boundaries. Let $\mathcal{R} \subseteq \mathcal{R}_{(D, \overline{D})}$ be the set of runs where all communication between the sets of processes $D_1, \ldots, D_{k-1}, \overline{D}$ is delayed until every correct process has decided, and assume that $\mathcal{R} \neq \emptyset$ (which will be proved in Lemma 12 below).

**Lemma 11.** *Let $\beta \in \mathcal{R}$ (and hence $\beta \in \mathcal{R}_{(D, \overline{D})}$ and $\beta \in \mathcal{R}_{(\overline{D})}$) and $\alpha \in \mathcal{R}_{(\overline{D})}$ be given, where $t^\alpha_{dec}$ resp. $t^\beta_{dec}$ denotes the time when the last process in $\overline{D}$ has crashed or decided in $\alpha$ resp. $\beta$. Then, the run $\beta'$ obtained from $\beta$ by*

1) *replacing $H_\beta(p, t)$ by $H_\alpha(p, t)$ at all processes $p \in \overline{D}$ at all times $t \geqslant 0$,*
2) *setting $F_{\beta'}(t) = (F_\beta(t) \cap (\Pi \setminus \overline{D})) \cup (F_\alpha(t) \cap \overline{D})$ at all times $t \geqslant 0$, and hence $F_{\beta'} = (F_\beta \cap (\Pi \setminus \overline{D})) \cup (F_\alpha \cap \overline{D})$,*
3) *letting the processes in $\overline{D}$ receive messages and perform their steps exactly as in $\alpha$,*
4) *delivering messages between $D_1, \ldots, D_k$ only after all correct processes have decided in $\beta'$,*
5) *choosing some (arbitratily large) $t_{GST} \geqslant \max\{t^\alpha_{dec}, t^\beta_{dec}\}$ and some set $LD$ that satisfies $LD \cap (\Pi \setminus F_{\beta'}) \neq \emptyset$, and setting $LD^t_j = LD$ in $H_{\beta'}$ for all processes $p_j \in \Pi \setminus F_{\beta'}$ and all $t \geqslant t_{GST}$*

*satisfies $\beta' \in \mathcal{R}$.*

*Proof:* We must show that (i) the processes in $D_1, \ldots, D_{k-1}$ see the same history in $\beta'$ and $\beta$ until $t_{GST}$, (ii) that the processes in $\overline{D}$ see the same history in $\beta'$ and $\alpha$ until $t_{GST}$, and (iii) that the history of $\beta'$ is a valid partitioning history: Then, correct processes can indeed take the same steps in $\beta'$ as in $\beta$ resp. $\alpha$, and hence decide the same. Since $\beta \in \mathcal{R}$ and $\alpha \in \mathcal{R}_{(\overline{D})}$, this implies $\beta' \in \mathcal{R}$.

Consider any $p_j \in D_i$, for some $1 \leqslant i \leqslant k$. The $\Sigma'_k$ output of processes in $D_i$ is not affected by changing the histories in $\overline{D}$, as the failure pattern for the processes in $D_i$ remains the same w.r.t. $\beta$ (for $1 \leqslant i \leqslant k-1$) resp. $\alpha$ (for $i = k$) and quorums in different partitions are disjoint according to Definition 7. The same holds true for the $\Omega'_k$ part, since the leader output of $p_j$ is independent of the leader output at any other process before stabilization time $t_{GST}$, after which it satisfies (Eventual Leadership) by construction. ∎

The next lemma shows that there are indeed "partitioned" executions, where all processes decide:

**Lemma 12.** $\mathcal{R}_{(D,\overline{D})} \neq \emptyset$, in particular, $R \subset \mathcal{R}_{(D,\overline{D})}$ is nonempty.

*Proof:* Fix any $D_i$, for $1 \leqslant i \leqslant k$ and $D_k = \overline{D}$. Consider a run $\alpha_i$ where all processes not in $D_i$ are initially dead; let $H_i$ denote the history of $\alpha_i$. Due to (Eventual Leadership) of $\Omega'_k$, there exists a time $t^i_{\text{GST}}$ and a set $LD^i$ with $LD^i \cap D_i \cap (\Pi \setminus F) \neq \emptyset$ such that $LD^t_j = LD^i$ for all $p_j \in D_i$ and $t \geqslant t^i_{\text{GST}}$. Since $A$ is correct, all correct processes in $D_i$ eventually decide in $\alpha_i$; let $t_i$ be the point in time when all processes in $D_i$ have either crashed or decided.

Using exactly the same arguments as in the proof of Lemma 11, we can construct a run $\alpha$ by "pasting" the executions $\alpha_i$, $1 \leqslant i \leqslant k$, one after the other: In the resulting $\alpha$, all processes in $\Pi$ fail exactly as in their respective $\alpha_i$, all communication between the sets $D_1, \ldots, D_k$ is delayed until time $\tau = \max(t_1, \ldots, t_k)$, and all processes in $D_i$, for every $1 \leqslant i \leqslant k$, take exactly the same steps as in $\alpha_i$. The history $H$ of $\alpha$ is equal to the union of the histories $H_i$ for all $t \geqslant 0$, except that we choose some (arbitratily large) $t_{\text{GST}} \geqslant \tau$ and any set $LD$ (which obviously satisfies $LD \cap D_j \cap (\Pi \setminus F) \neq \emptyset$ for some $j$) and require $LD^t_j = LD$ for all processes $p_j \in \Pi \setminus F$ and all $t \geqslant t_{\text{GST}}$. Obviously, $H$ satisfies Definition 7, so $\alpha$ is admissible and all processes decide in $\alpha$, i.e., $\alpha \in \mathcal{R}$. ∎

Equipped with these results, we can now establish the conditions required for applying Theorem 2:

**(A)** Consider the run $\alpha_k$ where all processes outside $\overline{D}$ are initially dead, then clearly processes in $\overline{D}$ decide before receiving a message from a processes outside $\overline{D}$. Since $\alpha_k \in \mathcal{R}_{(\overline{D})}$, we obviously have $\mathcal{R}_{(\overline{D})} \neq \emptyset$.

**(B)** Consider any run $\alpha \in \mathcal{R}_{(\overline{D})}$ with a (partitioning) failure detector history $H_\alpha$ and failure pattern $F_\alpha$, and let $\mathcal{R} \subseteq \mathcal{R}_{(D,\overline{D})}$ be the set of runs where all communication between the sets of processes $D_1, \ldots, D_{k-1}, \overline{D}$ is delayed until every correct process has decided. Now choose a run $\beta \in \mathcal{R}$ that satisfies the conditions of Lemma 11; Lemma 12 guarantees that it exists. Lemma 11 provides us with an execution $\beta' \in \mathcal{R}$ that is indistinguishable from $\alpha$ for all processes in $\overline{D}$ until decision, i.e., $\alpha \overset{\overline{D}}{\sim} \beta'$; recalling that $\beta' \in \mathcal{R} \subseteq \mathcal{R}_{(D,\overline{D})}$, we have thus shown that $\mathcal{R}_{(\overline{D})} \preccurlyeq_{\overline{D}} \mathcal{R}_{(D,\overline{D})}$.

**(C)** We will first choose an appropriately restricted model $\mathcal{M}'$: Since $|\overline{D}| = j \geqslant 3$, let $\mathcal{M}' = \langle \overline{D} \rangle$ be an asynchronous system where up to $j - 1$ processes may fail by crashing. Moreover, $\mathcal{M}'$ is augmented with a failure detector that is compatible to $(\Sigma'_k, \Omega'_k)$, in the sense that its failure detector histories can be extended to match an admissible history of $(\Sigma'_k, \Omega'_k)$ in $\mathcal{M}$, without changing the output at processes in $\overline{D}$: Considering Definition 7, we just assume that processes in $\mathcal{M}'$ effectively access a failure detector $(\Sigma, \Gamma)$, where $\Gamma$ satisfies the part of Definition 7 that concerns $\Omega_k$ in the following constrained way, for all processes in $\overline{D}$: $\Gamma$ outputs a possibly changing set of $k$ process ids in the range of $\Pi$, which eventually stabilizes on some set $LD$ that intersects $\overline{D}$ in exactly two processes $p_s$ and $p_t$. Obviously, this restriction is compatible with $\Omega'_k$. Note that one of $p_s$ and $p_t$ (but not necessarily both) may be faulty. Using $\Gamma$ we can easily implement $\Omega_2$ for $\mathcal{M}'$ (the transformation uses $\Gamma$ to eventually choose two fixed processes from $\overline{D}$), thus $(\Sigma, \Gamma)$ is weaker than $(\Sigma, \Omega_2)$. Moreover, $(\Sigma, \Omega_2)$ is strictly weaker than $(\Sigma, \Omega)$, as there is no transformation $T$ providing the properties of $(\Sigma, \Omega)$ from those of $(\Sigma, \Omega_2)$. If $T$ existed, we could use it to obtain a wait-free transformation $T'$ for shared memory to obtain $\Omega$ from $\Omega_2$ (by simulating an asynchronous message passing system equipped with $\Sigma$, cf. [9]) which contradicts the results of [21]. Since $(\Sigma, \Omega)$ is the weakest failure detector for solving consensus, we can therefore conclude that $(\Sigma, \Gamma)$ is too weak for solving consensus in $\mathcal{M}'$.

**(D)** Finally, for any run in $\mathcal{M}'_{A|\overline{D}}$, there is obviously a run in $\mathcal{R}_{(\overline{D})}$ where all processes in $D$ are initially dead, the processes in $\overline{D}$ take identical steps, fail at the same time, and receive the same failure detector output and the same messages. Hence, $\mathcal{M}'_{A|\overline{D}} \preccurlyeq_{\overline{D}} \mathcal{R}_{(\overline{D})}$ and, by transitivity, $\mathcal{M}'_{A|\overline{D}} \preccurlyeq_{\overline{D}} \mathcal{M}_A$. Applying Theorem 1 thus yields the required contradiction. ∎

From [3], we know that $\Sigma_{n-1}$ is sufficient for solving $(n-1)$-set agreement, Obviously, this implies that $(\Sigma_{n-1}, \Omega_{n-1})$ is also sufficient for $(n-1)$-set agreement. Together with the fact that $(\Sigma_1, \Omega_1)$ is sufficient for solving consensus [10], we have the following result:

**Corollary 13.** *There is an $(n-1)$-resilient algorithm that solves $k$-set agreement with failure detector class $(\Sigma_k, \Omega_k)_{1 \leqslant k \leqslant n-1}$ in an asynchronous system, if and only if $k = 1$ or $k = n-1$.*

## VIII. Discussion

In this paper, we introduced a reduction to consensus for generically characterizing the impossibility of $k$-set agreement in message passing systems. The main advantage of our approach is that we are independent of a specific system model, since Theorem 1 neither makes assumptions on the available amount of synchrony, nor on the power of computing steps and communication primitives available to the processes. This genericity allows to apply our theorem in very different contexts. In this paper, we have used our result to derive impossibility results both for partially synchronous systems [12] and for asynchronous systems augmented with failure detectors [3], [6]. However, we are confident that it can also be used to establish impossibility results in round models like [8], [15], [19].

A particularly promising application of our theorem is as both a guidance and quick verification tool for finding new models and algorithms for $k$-set agreement. This is particularly true for the quest for the (still unknown) weakest failure detector for solving message-passing $k$-set agreement: As we have shown, $\Sigma_k$, which is known to be necessary for $k$-set agreement in [3], is not powerful enough for overcoming the fatal partitioning into $k$ subsystems. So what can be learned from our result is that, whatever one adds to $\Sigma_k$, it has to allow solving consensus in each partition.

Our future work on this topic will involve (i) identifying other settings where Theorem 1 can be applied, (ii) developing a general theory of $T$-independence for failure detectors and other message passing systems, and (iii) finding weak system models that provide just enough synchrony to circumvent the impossibility condition.